\documentclass[a4paper,11pt]{article}
\usepackage{pos}
\usepackage{amsmath}

\usepackage{amssymb,graphicx,booktabs}
\usepackage{xcolor}

\providecommand{\ccbar}{c\overline{c}}
\providecommand{\bbbar}{b\overline{b}}
\providecommand{\ttbar}{t\overline{t}}
\newcommand{\sqrts}{\!\sqrt{s}}
\newcommand{\sqrtsnn}{\!\sqrt{s_{_\text{NN}}}}

\newcommand{\gag}{g_{a\gamma}}
\newcommand{\epem}{e^+e^-}
\newcommand{\gaga}{\gamma\gamma}

\title{The Future Circular Collider as a photon-photon collider}
\ShortTitle{The FCC as a photon collider}
\author*[a]{Patricia Rebello Teles}
\author[b]{David d'Enterria}
\affiliation[a]{Centro Brasileiro de Pesquisas F\'isicas (CBPF), Rio de Janeiro, Brazil}
\affiliation[b]{EP Department, CERN, Geneva, Switzerland}

\abstract{The CERN Future Circular Collider (FCC) offers a unique programme of photon–photon interactions in electron–positron, proton–proton, proton–lead, and lead–lead collisions, by exploiting the high-luminosity equivalent photon fluxes 
associated with the charged beams. Within the equivalent photon approximation, the electromagnetic fields of relativistic charged particles are described in terms of fluxes of quasireal photons. The complementary photon fluxes, energies, and virtualities available across the FCC operation modes provide access to a broad range of two-photon invariant masses and associated physics measurements. Physics opportunities open to study are highlighted, focusing on the exclusive production of $C$-even resonances, including the SM Higgs boson and searches for spin-0 axion-like (ALPs) and spin-2 graviton-like (GLPs) particles decaying back into photon pairs. These prospective studies demonstrate the potential to observe photon-fusion production of the Higgs boson for all colliding modes, as well as sensitivity to photon-coupled ALPs and GLPs with masses extending up to the kinematic reach of FCC-ee and into the multi-TeV range at FCC-hh collisions, with significant improvements over existing constraints.

}
\FullConference{43rd International Conference on High Energy Physics (ICHEP2026)\\
30 July -- 5 August 2026\\
Natal, Brazil}
\begin{document}
\maketitle
\section{The FCC as a multi-regime photon collider}
The Future Circular Collider (FCC) is a proposed CERN collider infrastructure based on a tunnel of $\sim$91~km circumference in the Geneva region, hosting first electron--positron (FCC-ee) followed by hadronic (FCC-hh) collisions~\cite{FCC:2025lpp}. The FCC-ee is planned to operate $\epem$ collisions at several centre-of-mass energies between the $Z$ pole and the $\ttbar$ threshold, $\sqrt{s}\approx 90$--365~GeV. The subsequent hadron collider will provide proton-proton ($pp$), proton-lead ($p$Pb), and lead-lead (PbPb) collisions at $\sqrtsnn \approx 39$--100~TeV. Although primarily designed for ultra-precise measurements of the Higgs, electroweak and top-quark properties, and for energy-frontier physics measurements, respectively, both machines also offer unique opportunities as photon-photon colliders, by exploiting the large electromagnetic (e.m.) fields associated with the relativistic charged beams.

A charged particle moving at relativistic speed carries an e.m.\ field that, in the equivalent photon approximation (EPA), can be represented by a spectrum of quasireal photons. The cross section for the exclusive production of a particle $X$ in photon-fusion collision of charged particles $A$ and $B$, which remain intact after the photon exchange, can be schematically written as the convolution of their photon fluxes $n_{A,B}(\omega_{1,2})$ with the underlying $\gaga\to X$ subprocess cross section,
\begin{equation}
\sigma(AB\to A+X+B)=\int \mathrm{d}\omega_1\,\mathrm{d}\omega_2\,n_A(\omega_1)n_B(\omega_2)\,\hat\sigma(\gaga\to X).
\label{eq:epa}
\end{equation}
This expression omits the additional dependence on photon virtuality, as well as impact-parameter effects and survival probabilities for hadron beams, which must be included for realistic predictions. 
At FCC-ee, the photon emitters are point-like electrons and positrons, whose QED photon spectra are perturbatively calculable~\cite{Budnev:1975poe}. In untagged two-photon events, the scattered beam leptons usually escape along the beam direction, while the central detector reconstructs the photon-fusion final state. This environment is particularly favourable for precision studies of photon-coupled particles. 

At FCC-hh, the photon spectra from protons and heavier ions depend on their electromagnetic form factors. In ultraperipheral collisions (UPCs), the impact parameter exceeds the sum of the colliding hadronic radii, suppressing their strong interaction and producing clean final states. Coherent photon emission from a nucleus with charge $Z$ gives a photon flux proportional to $Z^2$,  corresponding to an approximate $Z^4$ enhancement of the $\gaga$ luminosity in collisions of identical ions. This enhancement comes at the cost of a coherence cutoff on the maximum photon energy, set by the inverse nuclear radius ($1/R_\mathrm{A}$) in the projectile rest frame and boosted to the collider frame by the Lorentz factor  $\gamma_\mathrm{L}$, $\omega_{\gamma}^\mathrm{max}=\gamma_\mathrm{L}/R_\mathrm{A}$. Consequently, PbPb UPCs provide very large photon fluxes at low diphoton masses, $p$Pb collisions combine a strong nuclear photon flux with a high photon-energy reach, and $pp$ collisions access the highest two-photon invariant masses~\cite{Shao:2022cly}. These complementary collision modes probe a broad range of photon-induced physics processes.

\section{Photon-photon luminosities for even-spin particle production}
\begin{figure}[htbp]
\centering
\includegraphics[width=.45\linewidth]{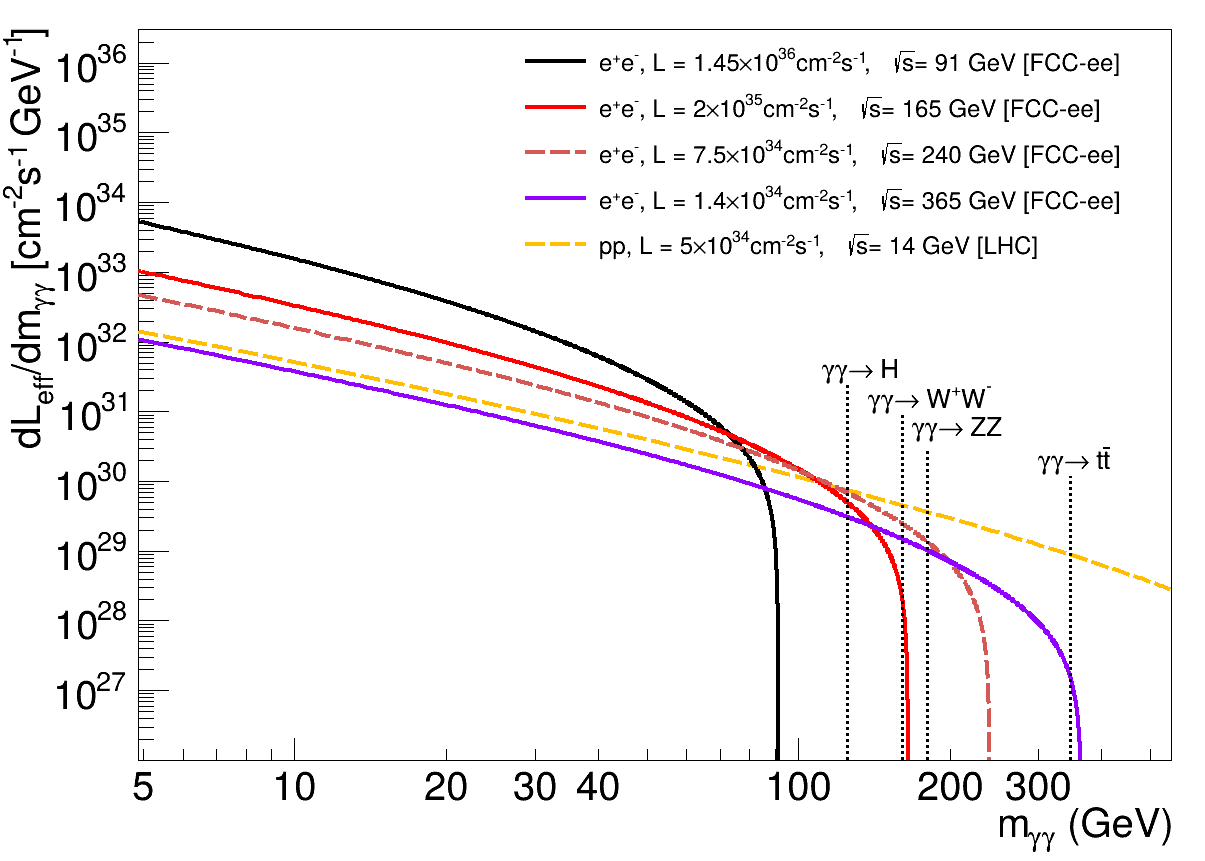}
\includegraphics[width=.53\linewidth]{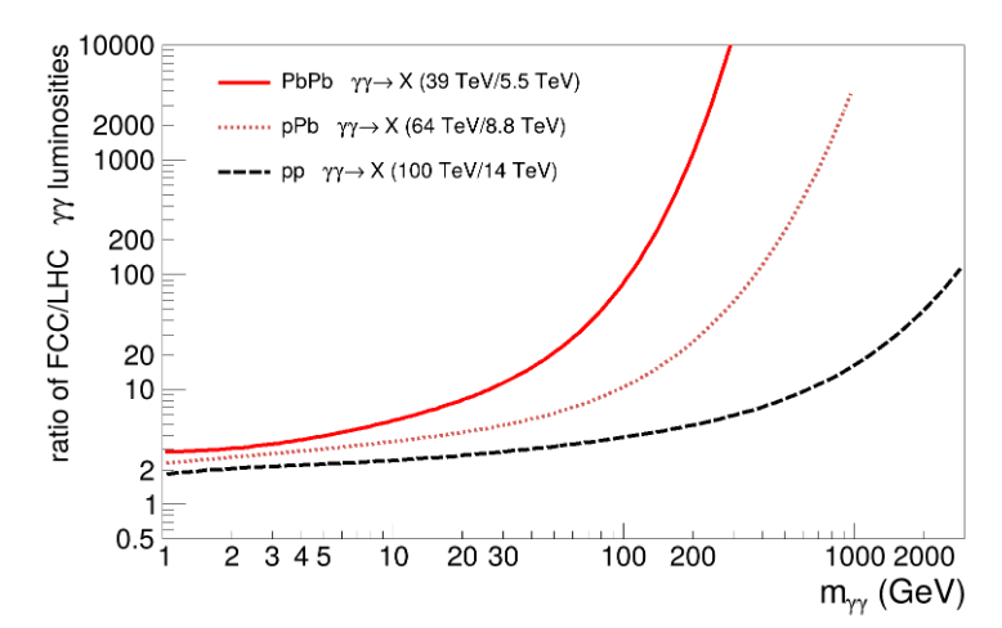}
\caption{
\textit{Left:} Effective instantaneous $\gamma\gamma$ luminosities as a function of $m_{\gaga}$, $dL_{\rm eff}/dm_{\gamma\gamma}$, in $\epem$ collisions at FCC-ee compared with those of $pp$ collisions at the LHC. The vertical markers indicate the kinematic thresholds for selected SM final states~\cite{dEnterria:2017jmj}.
\textit{Right:} Ratio of the FCC-hh/LHC effective $\gamma\gamma$ luminosities as a function of diphoton invariant mass $m_{\gamma\gamma}$, for PbPb, $p$Pb, and $pp$ collisions at the indicated c.m.\ energies. 
\label{fig:lum}}
\end{figure}

Two quasireal photons have positive charge-conjugation parity, $C(\gaga)=+1$, making photon fusion a selective probe of $C$-even resonances. For on-shell photons, the Landau--Yang theorem additionally forbids the production of a single spin-one state. The production of a $C$-even resonance $X$ of spin $J=0,2,...$, as considered here, in photon-fusion interactions of charged particles $A$ and $B$ 
reads~\cite{Budnev:1975poe},
\begin{equation}
\sigma(A B \xrightarrow{\gaga} A~X~B) = \left.4 \pi^2(2 J+1) \frac{\Gamma_{\gaga}}{m_{X}^2} \frac{\mathrm{d} \mathcal{L}_{\gaga}^{(\mathrm{AB})}}{\mathrm{d} m_{\gaga}}\right|_{m_{\gaga}=m_{X}},
\label{eq:sigma_X_master}
\end{equation}
where $\Gamma_{\gamma\gamma}$ denotes the partial decay width of $X$ into two photons, and $\left.\frac{\mathrm{d} \mathcal{L}_{\gaga}^{(\mathrm{AB})}}{\mathrm{d}m_{\gaga}}\right|_{m_{\gaga}=m_\mathrm{X}}$ is the photon-photon luminosity evaluated at the resonance mass $m_\mathrm{X}$.
Figure~\ref{fig:lum} (left) shows the large effective instantaneous diphoton luminosities achievable at FCC-ee thanks to the high beam luminosities of the machine. The right panel highlights the substantial increase in diphoton luminosities in UPCs from the LHC to the FCC-hh, in particular torwards the high-mass tail of the distributions.
The projected FCC/LHC enhancement is about 2--5 at $m_{\gaga}=1$--$10$~GeV, $10$--$100$ at $10$--$100$~GeV, and $10^2$--$10^4$ over $100$--$2000$~GeV. This increase follows the expected two-photon luminosity gain as a function of collision energy, $[\ln(\sqrtsnn/\sqrt{s_0}\,)/\ln(\sqrt{s'_{_\text{NN}}}/\sqrt{s'_0}\,)]^3$ where $\sqrt{s_0} = m_N R_\mathrm{A} m_{\gaga}$ is the minimum c.m.\ per nucleon pair required to produce a diphoton system of mass $m_{\gaga}$ without nuclear overlap ($2R_\mathrm{A}\approx 14$~fm for Pb)~\cite{dEnterria:2025ecx}. These estimates do not include the additional enhancement expected from the roughly tenfold increase in \emph{integrated} luminosities at FCC-hh.
The highest accessible diphoton masses extend well beyond those of the corresponding LHC running scenarios.

We have conducted feasibility studies of three types of $C$-even resonances produced in photon-photon collisions at the FCC: the Standard Model (SM) Higgs boson, via $\gaga\to H\to \bbbar$~\cite{dEnterria:2017jmj,dEnterria:2017yai,Higgs2020}, and beyond-the-Standard-Model (BSM) spin-0 axion-like-particles, $\gaga\to a\to\gaga$~\cite{ALPs2024}, and spin-2 graviton-like particles (GLPs), $\gaga\to G\to\gaga$~\cite{GLPs}. Light-by-light (LbL) scattering, $\gaga\to\gaga$, is both a fundamental QED process and an irreducible continuum background for the two latter resonance searches~\cite{dEnterria:2013zqi}. Other $C$-even SM resonances at FCC-hh are covered in Ref.~\cite{dEnterria:2025ecx}.

\section{Exclusive Higgs photon-fusion production at FCC-ee and FCC-hh}
The prospects for exclusive photon-fusion Higgs production, followed by its dominant decay ($\mathcal{B}\approx 58\%$) into a pair of b-quarks $\gaga\to H\to \bbbar$ have been studied at the FCC-ee~\cite{dEnterria:2017jmj} as well as in UPCs at FCC-hh with proton and ion beams~\cite{dEnterria:2017yai,Higgs2020}. This process probes the effective $H\gamma\gamma$ interaction through Higgs production rather than decay, which is the only method used to measure the coupling to date. In particular, at FCC-ee it provides direct sensitivity to the CP structure of the $H\gaga$ vertex through the azimuthal-angle correlation $\Delta\phi$ between the tagged outgoing leptons. The $\Delta\phi$ distribution retains information on the linear polarisation of the quasireal photons and is sensitive to any CP-odd admixture in the coupling, not accessible through the $\epem\to ZH$ measurement. The experimental signature comprises two exclusively reconstructed $b$-tagged jets with a pair mass peaked at $m_H\simeq125$~GeV 
on top of the $\gaga\to \bbbar$ background, with smaller additional contributions from $\gaga\to c\bar c, q\bar q$ events in which charm and light-flavour quarks are misidentified as $b$ quarks.

\begin{figure}[tbp]\centering
\includegraphics[width=.39\linewidth]{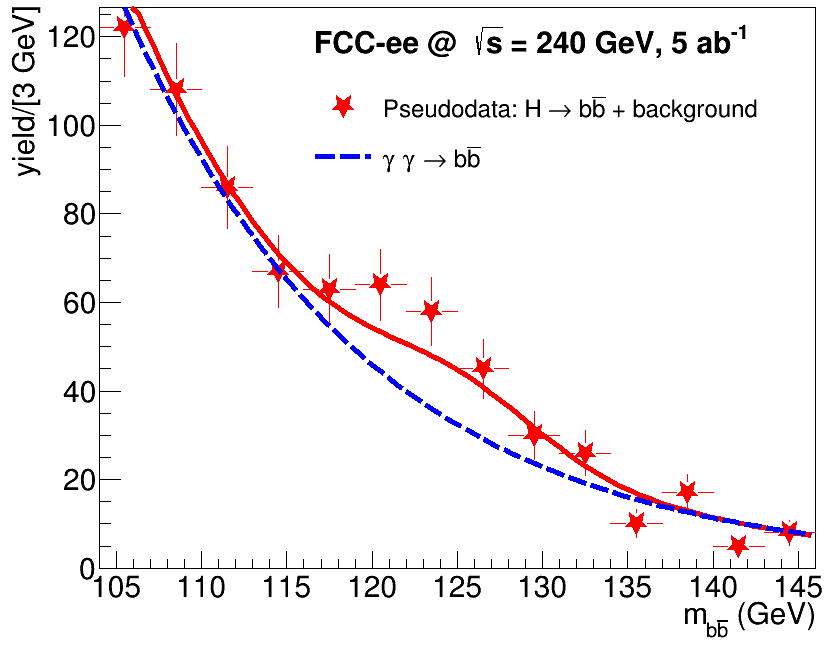}\hspace{5mm}
\includegraphics[width=.40\linewidth]{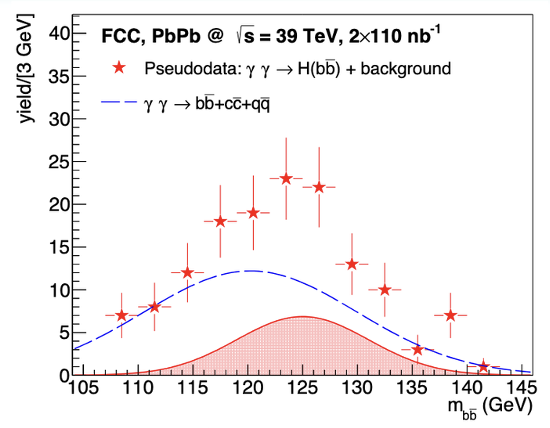}
\caption{Expected dijet invariant-mass distributions, after analysis cuts, for exclusive $\gamma\gamma \to H \to b\bar{b}$ production at FCC-ee at $\sqrt{s}=240$~GeV (left)~\cite{dEnterria:2017jmj}, and in PbPb UPCs at $\sqrt{s_{\mathrm{NN}}}=39$~TeV~(right)~\cite{Higgs2020}. The red stars show the expected signal-plus-background invariant mass counts. The Higgs~contribution is visible as an excess around $m_{b\bar{b}}\simeq125$~GeV on top of the $\gaga\to\ccbar$ continuum (dashed blue curves).
\label{fig:higgs}}
\end{figure}

With the latest FCC-ee integrated luminosities (19.2, 10.8,  $0.42+2.72$~ab$^{-1}$ for the WW, ZH, $\ttbar$ runs, respectively)~\cite{FCC:2025lpp}, the process $\gaga\to H\to \bbbar$ yields about 1600 counts across all FCC-ee centre-of-mass energies, corresponding to a combined statistical uncertainty of $\pm2.5\%$ on the rate before analysis cuts. 
The generator-level analysis performed in Ref.~\cite{dEnterria:2017jmj} indicated that the Higgs signal could be identified with a $5\sigma$ significance over the $\gaga\to \bbbar$ continuum using a simple event selection and 5~ab$^{-1}$ collected at $\sqrts = 240$~GeV (Fig.~\ref{fig:higgs} left). More advanced studies could significantly improve these results and provide constraints on the CP structure of the $H\gaga$ vertex. In heavy-ion UPCs at FCC-hh, the cross sections reach $\sigma(\gaga\to H) \approx 1.75$~nb and 1.5 pb in Pb-Pb and $p$Pb collisions at $\sqrtsnn = 39$ and $63$~TeV, respectively~\cite{Higgs2020}. These correspond to approximately 200 and 50 Higgs bosons for integrated luminosities of 110 nb$^{-1}$ and 29 pb$^{-1}$, respectively. A simple cut-based generator-level analysis similar to that performed for FCC-ee indicates about 20 $H(\bbbar)$ counts over a comparable number of background events in the 
$m(\bbbar) \approx 116$--134~GeV window, with a projected $5\sigma$ significance for
$2\times 110$~nb$^{-1}$ (Fig.~\ref{fig:higgs}, right)~\cite{Higgs2020}.


\section{Searches for axion-like particles}
\vspace{-6pt}
Axion-like particles (ALPs) are BSM pseudoscalar particles that can couple to photons with strength $\gag$ through an effective interaction of the form
\begin{equation}
\mathcal{L}_{a\gamma\gamma}=-\frac14\gag\,a F_{\mu\nu}\widetilde F^{\mu\nu}.
\label{eq:alp}
\end{equation}
Typically, searches assume a photophilic scenario in which the $a\to\gaga$ decay dominates. The $\gaga\to a\to \gaga$ production and decay process is
characterized by a diphoton invariant-mass peak over the exclusive LbL scattering continuum. In the narrow-width regime, the sensitivity is conventionally expressed in the $(m_a,\gag)$ plane. 
The FCC-ee projections use the \textsc{SuperChic}~v4 event generator~\cite{Harland-Lang:2020veo} and a fast \textsc{Delphes} detector simulation based on the IDEA concept. The FCC-hh studies use \textsc{gamma-UPC}~\cite{Shao:2022cly} interfaced with \textsc{MadGraph5\_aMC@NLO}, 
and the \textsc{Delphes} detector card. The hadronic-collider analysis selects two well-reconstructed photons with no additional activity associated with the interaction vertex and a nearly back-to-back topology characterized by a diphoton acoplanarity below 1\%. The stated detector performances consider efficient photon reconstruction down to transverse momenta of a few GeV over broad pseudorapidity range.

\begin{figure}[tbp]\centering
\includegraphics[width=.49\linewidth]{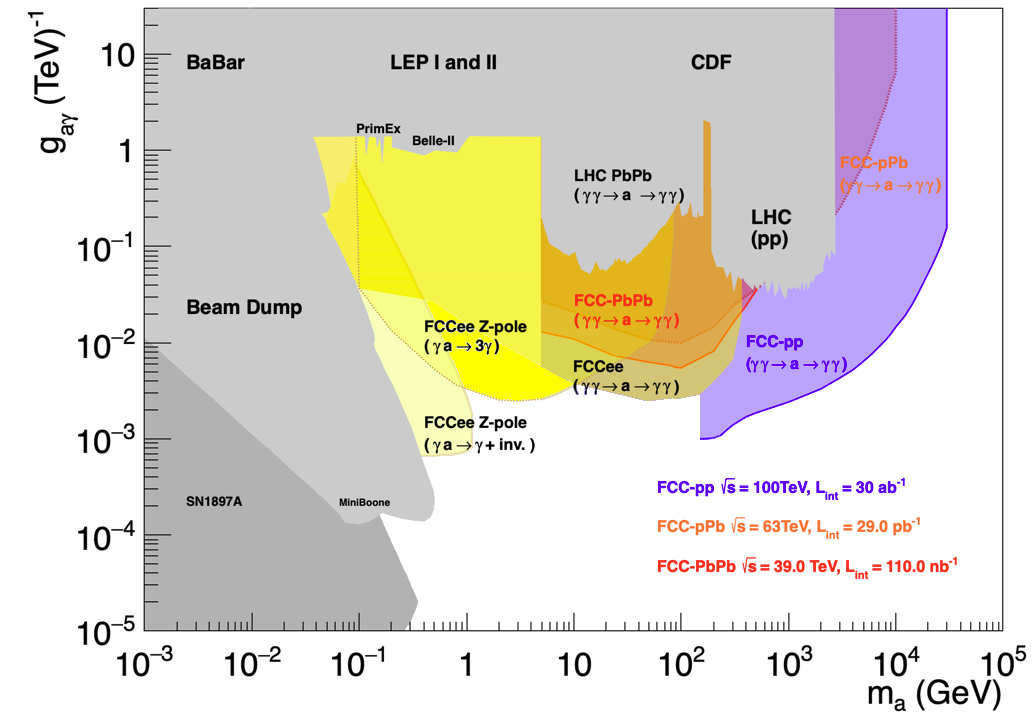}
\includegraphics[width=.50\linewidth]{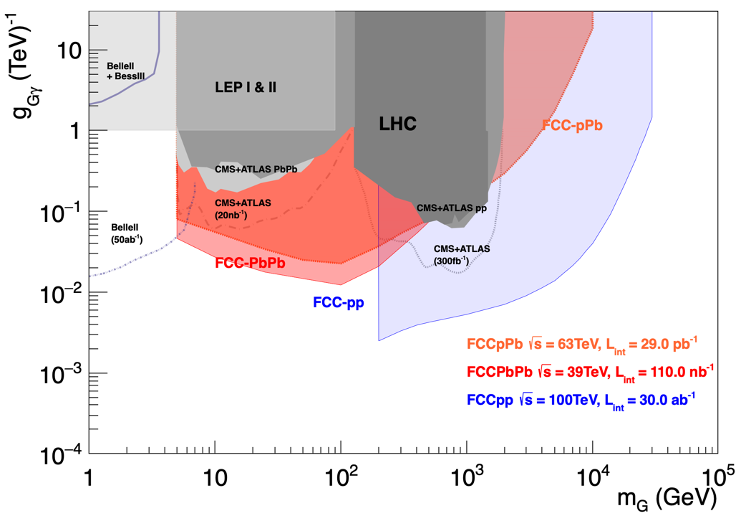}
\caption{Projected ALP (left) and GLP (right) sensitivity in $\gaga$ collisions at different FCC running modes (colored regions) in the photon-coupling vs.\ ALP/GLP mass, compared to current limits (gray regions)~\cite{ALPs2024,GLPs}. 
\label{fig:alp_glp}}
\end{figure}

The FCC-ee can probe ALP masses approaching the kinematic reach of its highest-energy stage, with coupling sensitivity as low as $\gag\approx 2\cdot 10^{-3}\,\mathrm{TeV}^{-1}$ over $m_a \approx 1$--350~GeV. The ALP diphoton width is proportional to $\Gamma(a\to\gamma\gamma)\propto \gag^2 m_a^3$ and, thus, light and/or weakly coupled ALPs may decay with a measurable displacement or outside the detector, invalidating a purely prompt-diphoton search assumption. Photon pointing, calorimeter granularity, and dedicated displaced-decay reconstruction therefore become important for $m_a\lesssim 1$~GeV~\cite{ALPs2024}. At FCC-hh, PbPb and $p$Pb UPCs are projected to improve the HL-LHC constraints by roughly one order of magnitude, while the $pp$ mode reaches masses of order $m_a\approx 30$~TeV and coupling sensitivities down to $10^{-3}\,\mathrm{TeV}^{-1}$ in the $m_a$ region of a few hundred GeV~(Fig.~\ref{fig:alp_glp}, left).
\section{Massive graviton-like resonances} 

Massive spin-2 graviton-like particles (GLPs) provide another benchmark for exclusive BSM particle production via photon fusion. Their phenomenological description is based on a massive spin-2 Pauli--Fierz effective theory supplemented by universal couplings to the SM energy-momentum tensors. A GLP produced by photon fusion can decay into two photons, giving rise to a resonance over the LbL scattering continuum. The spin factor $(2J+1)$ in Eq.~(\ref{eq:sigma_X_master}) enhances production relative to ALPs for otherwise comparable couplings, although a direct comparison must account for their different partial widths~\cite{Gue:2026wqd}. In the universal-coupling benchmark, additional SM decay modes reduce the diphoton branching fraction to a few percent, with $\mathcal{B}(G\to\gaga )\approx 5$--10\% for $m_G\gtrsim 1$~GeV. The GLP projections for FCC-hh use the same simulation chain as the ALP analysis, including \textsc{gamma-UPC} EPA fluxes, matrix-element generation, particle decays, and fast detector simulation~\cite{GLPs}. At FCC-hh , the projected $pp$ sensitivity improves on current LHC bounds by about one to two orders of magnitude for $m_G\simeq0.2$--1~TeV, and extends to previously unexplored masses of order 2--30~TeV, whereas PbPb and \(p\)Pb UPCs provide complementary sensitivity at lower masses, including the $m_G\approx 5$--100~GeV  range (Fig.~\ref{fig:alp_glp}, right).

\section{Conclusions}

The FCC can be viewed as a multi-regime photon collider, with its $\epem$, $pp$, $p$Pb, and PbPb operating modes providing complementary photon fluxes and access to a broad range of two-photon invariant masses. Our studies have focused on the exclusive production of $C$-even SM and BSM resonances through photon–photon fusion.
First, FCC-ee and FCC-hh offer the opportunity to measure exclusive  $\gaga\to H$ production for the first time and to probe the effective Higgs–photon coupling in detail. Second, searches for ALPs and massive GLPs exploit the same production mechanism to probe BSM spin-0 and spin-2 resonances over a mass range extending from the GeV scale to tens of TeV. At FCC-hh, forward proton tagging could provide an additional handle for selecting exclusive events, although its acceptance and feasibility at multi-TeV invariant masses require dedicated study. Further work will focus on full detector simulations, more detailed background studies, the development and optimisation of advanced event-selection strategies, and a comprehensive assessment of systematic uncertainties.

\vspace{-6pt}
\acknowledgments
\vspace{-6pt}\hspace{-0.7cm}Patricia Rebello Teles warmly thanks the support of CNPq via the grant 201440/2026-1.

\vspace{-6pt}
{\footnotesize\setlength{\parskip}{2pt}
}
\end{document}